\documentclass[prd,twocolumn,aps,superscriptaddress,nofootinbib]{revtex4-1}
\usepackage{latexsym,amssymb,amsmath,epsfig,bm,psfrag}
\usepackage{color}
\usepackage[bookmarksnumbered,bookmarksopen,colorlinks,citecolor=red,linkcolor=blue]{hyperref}
\usepackage{mathrsfs}
\usepackage{times}
\newcommand\be{\begin{equation}}
\newcommand\ba{\begin{eqnarray}}
\newcommand\ee{\end{equation}}
\newcommand\ea{\end{eqnarray}}

\hypersetup{
    colorlinks=true,
    linkcolor=blue,
    filecolor=magenta,
    urlcolor=cyan,
    pdftitle={Sharelatex Example},
    bookmarks=true,
    pdfpagemode=FullScreen,
}

\newcommand{\hzero}{{\hat{0}}}

\newcommand{\lD}{\overleftarrow{D}}

\newcommand{\calV}{\mathcal{V}}
\newcommand{\calA}{\mathcal{A}}
\newcommand{\calS}{\mathcal{S}}
\newcommand{\calR}{\mathcal{R}}
\newcommand{\calL}{\mathcal{L}}
\newcommand{\calW}{\mathcal{W}}

\newcommand{\bh}{{\boldsymbol \theta}}
\newcommand{\dis}{\displaystyle}
\newcommand{\tr}{\text{tr}}
\newcommand{\zero}{ {(0)} }
\newcommand{\one}{ {(1)} }
\newcommand{\bzero}{{\boldsymbol 0}}
\newcommand{\bp}{{\boldsymbol p}}

\newcommand{\LE}{\text{LE}}
\newcommand{\GE}{\text{GE}}

\renewcommand\a{\alpha}
\renewcommand\b{\beta}
\renewcommand\d{\delta}

\renewcommand\l{\lambda}
\renewcommand\r{\rho}

\renewcommand\t{\tau}

\renewcommand\j{\psi}
\renewcommand\o{\omega}
\newcommand\e{\epsilon}
\newcommand\g{\gamma}

\newcommand\m{\mu}
\newcommand\n{\nu}

\newcommand\p{\pi}
\newcommand\h{\theta}
\newcommand\s{\sigma}
\newcommand\f{\phi}
\newcommand\w{\eta}

\renewcommand\L{\Lambda}
\renewcommand\P{\Pi}
\renewcommand\S{\Sigma}

\newcommand\D{\Delta}
\newcommand\G{\Gamma}

\newcommand\lb{\left(}
\newcommand\rb{\right)}
\newcommand\ls{\left[}
\newcommand\rs{\right]}

\newcommand{\non}{\nonumber\\}
\newcommand\pt{\partial}
\newcommand\mc{\mathcal}
\newcommand\ms{\mathscr}
\newcommand\na{\nabla}
\newcommand\ola{\overleftarrow}

\newcommand{\Tr}{{\rm Tr}}

\renewcommand{\part}{{\rm part}}

\begin{document}

\title{Covariant Spin Kinetic Theory I: Collisionless Limit}
\author{Yu-Chen~Liu}
\email{ycliu11@fudan.edu.cn}
\affiliation{Physics Department and Center for Particle Physics and Field Theory, Fudan University, \\ Shanghai 200433, China}
\author{Kazuya~Mameda}
\email{kazuya.mameda@riken.jp}
\affiliation{Physics Department and Center for Particle Physics and Field Theory, Fudan University, \\ Shanghai 200433, China}
\affiliation{RIKEN Nishina Center for Accelerator-Based Science, Wako 351-0198, Japan}
\author{Xu-Guang~Huang}
\email{huangxuguang@fudan.edu.cn}
\affiliation{Physics Department and Center for Particle Physics and Field Theory, Fudan University, \\ Shanghai 200433, China}
\affiliation{Key Laboratory of Nuclear Physics and Ion-beam Application (MOE), Fudan University, \\ Shanghai 200433, China}

\begin{abstract}
We develop a covariant kinetic theory for massive fermions in a curved spacetime and an external electromagnetic field based on quantum field theory.
We derive four coupled semi-classical kinetic equations accurate to $O(\hbar)$, which describe the transports of particle number and spin degrees of freedom.
The relationship with chiral kinetic theory is discussed.
As an application, we study spin polarization in the presence of finite Riemann curvature and an electromagnetic field in both local and global equilibrium states.\\
\vspace{0pt}
\end{abstract}
\maketitle

\section{Introduction}
Kinetic theory is widely used to study transport phenomena in many-particle systems.
The classical Boltzmann kinetic theory has been established as the framework to describe the evolution of the distribution function in phase space.
To study the effects of spin, the quantum kinetic theory must be used, and a single distribution function is usually insufficient for such a purpose.
For massless Dirac fermions, the leading spin effects appear at $O(\hbar)$; thus, two distribution functions are required, one for right-handed chirality and the other for left-handed chirality%
~\footnote{%
A massless particle and antiparticle with its spin parallel (anti-parallel) to its momentum is called to have right-(left-)handed and the left-(right-)handed chirality, respectively.%
}.
This established framework is the chiral kinetic theory (CKT)~\cite{Son:2012wh,Stephanov:2012ki,Gao:2012ix}, which has been intensively investigated recently~\cite{Son:2012zy,Chen:2012ca,Chen:2014cla,Chen:2015gta,Hidaka:2016yjf,Mueller:2017arw,Carignano:2018gqt,Huang:2018wdl,Huang:2018aly,Liu:2018xip,Lin:2019ytz}.
The out-of-equilibrium dynamics of anomaly-induced phenomena, such as the chiral magnetic effect~\cite{Kharzeev:2007jp,Fukushima:2008xe} and the chiral vortical effect~\cite{Vilenkin:1979ui,Erdmenger:2008rm,Banerjee:2008th}, have also been thoroughly investigated in the framework of CKT.

Unlike the massless case, the spin of a massive fermion is independent of the kinetic momentum.
As a result, the dynamical evolution of massive fermions is specified with more than two degrees of freedom. If $\theta^\m$ is the unit space-like vector specifying the spin quantizing orientation,
the dynamical variables are two parameters used to determine $\theta^\m$, and two distribution functions, $f_\pm$, for particles with spin parallel and anti-parallel to $\theta^\m$, respectively.
Therefore, the kinetic theory of massive fermions is more complicated than the CKT, and extensive investigation of this framework is needed~\cite{Gao:2019znl,Weickgenannt:2019dks,Hattori:2019ahi,Wang:2019moi}.

One of the motivations to develop the aforementioned kinetic theory is the spin-polarization phenomenon in heavy-ion collisions, which is an important probe of the hot and dense quark gluon matter~\cite{Liang:2004ph,Gao:2007bc,Becattini:2007sr,Huang:2011ru}.
The first signal of the global spin polarization of $\L$ hyperons (hereafter, $\L$ polarization)~\cite{STAR:2017ckg} indicates the existence of a very strong fluid vorticity~\cite{Deng:2016gyh,Jiang:2016woz,Deng:2020ygd}.
The subsequent measurements exhibit very nontrivial features that cannot be understood based on the simple vorticity interpretation of the spin polarization~\cite{Adam:2018ivw,Adam:2019srw}.
For example, the measured longitudinal and transverse $\L$ polarization show the opposite azimuthal angle dependence compared with the so-called thermal vorticity~\cite{Karpenko:2016jyx,Becattini:2017gcx,Xia:2018tes,Sun:2018bjl,Wei:2018zfb,Xia:2019fjf,Becattini:2019ntv,Wu:2019eyi}.
This strongly indicates that the spin polarization has independent dynamical evolution, in a non-equilibrium state, rather than being chained to the fluid vorticity. A covariant kinetic theory for spin transport (hereafter, {\it spin kinetic theory} for short) would be a promising tool to capture the dynamics of the spin polarization.

In this paper, we derive the collisionless spin kinetic theory at $O(\hbar)$ in a curved background spacetime and an external electromagnetic field.
As an application, we investigate the spin polarization using the spin kinetic theory. We give the general expression for spin polarization in terms of $f_\pm$ and $\h^\m$. We then specify the equilibrium conditions from the spin kinetic theory and derive the spin polarization at both local and global equilibria.
We stress that the present study differs from the earlier works~\cite{Gao:2019znl,Weickgenannt:2019dks,Hattori:2019ahi,Wang:2019moi} in the following aspects. 
First, we include curved geometric background spacetime as well as an electromagnetic field.
Such a general formalism should be applicable to the spin transport, not only in heavy-ion collisions and astrophysical systems, but also in deformed materials and thermal-gradient systems, which attracts significant attention in condensed matter physics. 
Second, we show that the frame-choosing vector can always be eliminated in the covariant kinetic theory of the massive fermions, unlike the massless case.
Third, we discuss the underlying physics of the Clifford components and their constraint equations.
Fourth, we provide the kinetic equations in a more transparent way, exhibiting clear physical contexts.
In particular, we verify that, in the classical limit, these equations are correctly reduced to the Vlasov equation, the Bargmann--Michel--Telegdi (BMT) equation~\cite{Bargmann:1959gz}, and Mathisson--Papapetrou--Dixon (MPD) equations~\cite{Mathisson:1937zz,Dixon:1970zza,Papapetrou:1951pa}.
Finally, we discuss the global equilibrium in terms of the spin vector $\theta^\m$.
The validity of this equilibrium state is qualified by the resulting spin polarization, which is consistent with that in Refs.~\cite{Becattini:2013fla,Fang:2016vpj}.

This paper is organized as follows: In Sections~\ref{sec:wigner} and~\ref{sec:phys}, we introduce the Wigner function and discuss the physical meaning of the dynamical equation for each Clifford component of the Wigner function. In Section~\ref{sec:qkt}, we derive the semi-classical kinetic theory for massive fermions. In Section~\ref{sec:sppol}, we derive the kinetic representation of the spin polarization for both massive and massless fermions and investigate the spin polarization at both local and global equilibria. In this paper, we adopt the same notations and conventions as those of Ref.~\cite{Liu:2018xip};
for instance, $\na_\m$ denotes the covariant derivative in terms of diffeomorphism and the local Lorentz transformation, and $p_\m$ ($y^\m$) is the momentum variable (its conjugate one).

\section{Wigner function}\label{sec:wigner}

The Wigner operator covariant under the $\text{U}(1)$ gauge, local Lorentz transformations, and diffeomorphism is defined as~\cite{Liu:2018xip}
\begin{eqnarray}
\label{wp}
& \dis
	\hat{W}(x,p)=\int d^4y\sqrt{-g(x)}\, e^{-ip\cdot y/\hbar} \hat{\r}(x,y) \,,\\
& \dis
	\hat{\r}(x,y) = \bar\psi(x)e^{y\cdot\ola{D}/2} \otimes e^{-y \cdot D/2}\j(x) \,,
\end{eqnarray}
where $\j(x)$ is the Dirac spinor operator.
Here we introduce the following notations: $\bar\psi(x)\equiv \psi^\dag(x)\gamma^\hzero$ and $\bar\psi\ola{O}\equiv [O \psi]^\dag \gamma^\hzero$ for an operator $O$, and $[\bar\psi\otimes \psi]_{ab}=\bar\psi_b \psi_a$ with $a,b$ being the spinor indices.
The derivative $D_\mu$ is called the horizontal lift of $\nabla_\mu$\,:
$D_\m = \na_\m - \G^{\l}_{\m\n}y^\n \pt_\l^y$ in the tangent bundle [i.e., the $(x,y)$-space]. Similarly the horizontal lift in the cotangent bundle [i.e., the $(x,p)$-space] is given by
\begin{equation}
 D_\m = \na_\m + \G^{\l}_{\m\n}p_\l \pt^\n_p \,.
\end{equation}
This $D_\mu$ gives us a great advantage regarding analysis because of the property $[D_\mu,y^\nu]=[D_\mu,p_\nu] = 0$.
We note that the gauge field $A_\m$ should also be involved in $D_\m$ when acting on the Dirac spinor: $D_\m\j(x,y)=(\na_\m-\G^{\l}_{\m\n}y^\n \pt_\l^y+iA_\m/\hbar)\j(x,y)$ with $\psi(x,y)\equiv e^{y\cdot D}\psi(x)$.

The Wigner function is defined by replacing the operator $\hat{\r}(x,y)$ with the ensemble average $\r(x,y)\equiv\langle\hat{\r}(x,y)\rangle$ in Eq.~\eqref{wp}.
In this paper, we focus on the collisionless limit; thus, we impose the spinor field to obey the Dirac equation $(i\hbar\gamma^\mu D_\mu-m)\psi(x)=\bar{\psi}(x)\,(i\hbar{\ola D}_\mu\gamma^\mu +m)=0$.
In this case, we derive the kinetic theory of massive fermions in the same manner as that in Ref.~\cite{Liu:2018xip} (in particular, see Section III and Appendices C, D, and E therein).
After the semi-classical expansion%
~\footnote{%
We employ the power counting scheme as $p_\m=O(1)$ and $y^\m\sim i\hbar\pt^\m_p=O(\hbar)$.%
}, and the decomposition in terms of the Clifford algebra as $W=\frac{1}{4}[\mathcal{F}+i\g^5\mathcal{P}+\g^\m\mathcal{V}_\m +\g^5\g^\m\mathcal{A}_\m+\frac{1}{2}\s^{\m\n}\mathcal{S}_{\m\n}]$, we arrive at the following system of equations:
\begin{eqnarray}
\label{eq:VWI}
& \dis
\D_\m \mc{V}^\m
 =
	\frac{\hbar^2}{24}(\na_\mu R_{\nu\rho})\pt^\mu_p\pt^\nu_p\mc{V}^\rho \,, \\
\label{eq:AWI}
& \dis
\hbar\D_\m \mc{A}^\m  =  - 2m\mc{P} \,, \\
\label{eq:Lorentz}
& \dis
\hbar\D_{[\m} \mc{A}_{\n]}
-\e_{\m\n\r\s}{\P}^\r\mc{V}^\s
 = -\frac{\hbar^2}{8}\tilde{R}_{\rho\sigma\mu\nu}\partial_p^\rho \calV^\sigma \,, \\
\label{eq:dilatation}
& \dis
{\P}_\m \mc{V}^\m = m\mc{F} + \frac{\hbar^2}{8}R_{\m\n}\pt^\m_p\mc{V}^\n \,, \\
\label{eq:pdotA}
& \dis
{\P}_\m \mc{A}^\m = \frac{\hbar^2}{8}R_{\m\n}\pt^\m_p\mc{A}^\n \,, \\
\label{eq:aLorentz}
& \dis
\hbar\D_{[\m} \mc{V}_{\n]}
-\e_{\m\n\r\s}{\P}^\r\mc{A}^\s
 = m\mc{S}_{\m\n}
 - \frac{\hbar^2}{8} \tilde{R}_{\rho\sigma\mu\nu} \partial^\rho_p \calA^\sigma \,, \\
\label{eq:GordonV}
& \dis
{\P}_\m \mc{F}+\frac{\hbar}{2}\D^\n\mc{S}_{\m\nu} = m \mc{V}_\m \,, \\
\label{eq:GordonA}
& \dis
\frac{\hbar}{2}\D_\m \mc{P}- {\P}^\n \tilde{\mc{S}}_{\mu\nu}
 = m\mc{A}_\m
	+ \frac{\hbar^2}{8} \tilde{R}_{\nu\rho\sigma\mu}\partial_p^\nu \calS^{\rho\sigma} \,, \\
\label{eq:GordonV_mag}
& \dis
{\P}_\m \mc{P}
+ \frac{\hbar}{2}\Delta^\nu \tilde\calS_{\mu\nu}
 = 0 \,, \\
\label{eq:GordonA_mag}
& \dis
\frac{\hbar}{2}\D_\m\mc{F}-{\P}^\n \mc{S}_{\m\n}
 = \frac{\hbar^2}{16}
	(R_{\m\n\r\sigma}\pt^\n_p \mc{S}^{\r\sigma}
		+ 2 R^{\nu\rho}\pt^p_\n \mc{S}_{\r\m}) \quad\ \ \
\ea
with $X_{[\mu}Y_{\nu]} = \frac{1}{2}(X_\mu Y_\nu - X_\nu Y_\mu)$.
Here ${R^\rho}_{\sigma\mu\nu} = 2\partial_{[\nu}\Gamma_{\mu]\sigma}^\rho +2\Gamma_{\lambda[\nu}^\rho\Gamma^\lambda_{\mu]\sigma}$ is the Riemann tensor, $R_{\mu\nu}={R^\rho}_{\mu\rho\nu}$ is the Ricci tensor, and we define $\tilde{\calS}^{\mu\nu} = \frac{1}{2}\epsilon^{\mu\nu\rho\sigma}\calS_{\rho\sigma}$ and $\tilde{R}^{\mu\nu\alpha\beta} = \frac{1}{2}\epsilon^{\alpha\beta\rho\sigma}{R^{\mu\nu}}_{\rho\sigma}$.
The operators $\Pi_\mu$ and $\Delta_\mu$ are given by
\begin{equation}
 \begin{split}
 \label{ptmd}
 {\P}_\mu
  &= p_\mu
   	- \frac{\hbar^2}{12}(\nabla_\rho F_{\mu\nu})\partial^\n_p\partial^\rho_p
  	  + \frac{\hbar^2}{24}{R^\rho}_{\sigma\mu\nu}\partial^\sigma_p\partial_p^\nu p_\rho \\
  & \quad
  	  +\frac{\hbar^2}{4}R_{\mu\nu}\partial_p^\nu, \\
 \Delta_\mu
  & = D_\mu -F_{\mu\lambda}\partial_p^\lambda
  	 - \frac{\hbar^2}{12}(\nabla_\r R_{\m\n})\partial_p^\r\partial_p^\nu \\
  &\quad  - \frac{\hbar^2}{24}(\nabla_\l{R^\rho}_{\sigma\mu\nu})
  	    \partial_p^\nu\partial_p^\s\partial_p^\lambda p_\rho
  	 + \frac{\hbar^2}{8}{R^\rho}_{\sigma\mu\nu}\partial_p^\nu\partial_p^\sigma D_\rho \\
  &\quad + \frac{\hbar^2}{24} ( \nabla_\alpha\nabla_\beta F_{\mu\nu}
  	  + 2{R^\rho}_{\alpha\mu\nu}F_{\beta\rho} )
  	    \partial_p^\nu \partial_p^\alpha\partial_p^\beta \,.
 \end{split}
\end{equation}
In equations~\eqref{eq:VWI}-\eqref{eq:GordonA_mag}, the spacetime curvature enters at least at $O(\hbar^2)$.
We note that the Clifford coefficients $\mathcal{F}$, $\mathcal{P}$, $\mathcal{V}_\m$, $\mathcal{A}_\m$, and $\mathcal{S}_{\m\n}$ are not totally independent.
To proceed, we choose $\mathcal{V}_\m$ and $\mathcal{A}_\m$ as the independent variables~\footnote{In fact, only two (for massless case) or four (for massive case) components of $(\mathcal{V}_\m, \mathcal{A}_\m)$ are independent.}.
Then $\mathcal{P}$, $\mathcal{F}$, and $\mathcal{S}_{\m\n}$ are expressed in terms of $\mathcal{V}_\m$ and $\mathcal{A}_\m$ through Eqs.~\eqref{eq:AWI},~\eqref{eq:dilatation}, and \eqref{eq:aLorentz}.
In Minkowski spacetime, the same set of equations up to $O(\hbar)$ was first derived in Ref.~\cite{Vasak:1987um}.

In the kinetic description, various physical quantities are built from $W$, which is (the Wigner transformation of) a two-point correlator of Dirac fields.
For instance, the vector and axial current are computed as
\begin{eqnarray}
 & \dis
 J^\mu = \int_p \tr\Bigl[\gamma^\mu W\Bigr]
 = \int_p \calV^\mu \,, \\
 & \dis
 J_{5}^\mu = \int_p \tr\Bigl[\gamma^\mu \gamma^5 W\Bigr]
 = \int_p \calA^\mu
\end{eqnarray}
with $\int_p \equiv\int \frac{d^4 p}{(2\p)^4 [-g(x)]^{1/2}}$.
From these, the Clifford coefficients $\mathcal{V}_\m$ and $\mathcal{A}_\m$ can be identified as the corresponding current densities in phase space (see more discussions in Sec.~\ref{sec:sppol}).
In a similar way, $\mathcal{F}$ is the scalar condensate density (which in the classical limit is also interpreted as the distribution function of the vector charge); $\mathcal{P}$ is the axial condensate density; and $\mathcal{S}_{\m\n}$ is the electromagnetic dipole moment density, up to a factor of $m$ [see Eqs.~\eqref{solutionf}-\eqref{solutions}].
For convenience, we further represent the canonical energy-momentum tensor, spin current, and total angular momentum current as
\begin{eqnarray}
 &\dis
 \label{eq:T}
 T^{\mu\nu}
 = \int_p
 	\tr\biggl[
 	\frac{i\hbar}{2}\gamma^\mu \overset{\leftrightarrow\;\;}{D^\nu} W
 	\biggr]
 = \int_p \mc{V}^\m p^\n \,,\\
 \label{eq:Sc}
 & \dis
 S^{\mu,\nu\rho}
 = \int_p
 	\tr\biggl[
 	\frac{\hbar}{4}\{\gamma^\mu,\sigma^{\nu\rho}\} W
 	\biggr]
 = -\frac{\hbar}{2}\int_p\e^{\mu\nu\rho\lambda}\mc{A}_\l \,, \\
 & \dis
 \label{eq:AMT}
 M^{\lambda,\mu\nu}
 = x^\mu T^{\lambda\nu} - x^\nu T^{\lambda\mu} + S^{\lambda,\mu\nu} \,,
\end{eqnarray}
where we define $\overset{\leftrightarrow}{D}_\mu (\bar\psi_b \otimes \psi_a) \equiv \bar\psi_b \otimes D_\mu \psi_a - \bar\psi_b \lD_\mu \otimes  \psi_a$.

Note that $S^{\lambda,\mu\nu}$ is not anticipated to be an observable for spin because there exists the Belinfante--Rosenfeld type pseudo-gauge ambiguity~\cite{Becattini:2018duy};
$S^{\lambda,\mu\nu}$ in Eq.~\eqref{eq:AMT} can be absorbed into a redefinition of the energy momentum tensor once the Belinfante tensor has been introduced.
Instead, an unambiguous way to represent particles spin is to employ the Pauli-Lubanski (PL) vector operator:
\begin{equation}
\label{eq:PLqm}
\hat{\calW}^{\m}
 \equiv
	-\frac{1}{\hbar}\e^{\m\n\r\s}\hat{P}_\n \hat{M}_{\r\s} \,,
\end{equation}
where the hat symbols denote quantum mechanical operators, $\hat{P}_\n$ and $\hat{M}_{\r\s}$ are the canonical momentum and total angular momentum operators, respectively, and the prefactor $-1/\hbar$ is introduced as our convention.
It is important to notice that the orbital part of the canonical angular momentum does not contribute to the above equation.

Following Eq.~\eqref{eq:PLqm}, we may define the PL vector in our kinetic theory as~\cite{Maybee:2019jus}%
~\footnote{%
This is not equivalent to the ensemble average of the PL vector operator $\hat{\calW}^\mu$.
Nevertheless, the resulting form~\eqref{eq:W-A} is the same as the one evaluated in the usual quantum field theory therein.}
\begin{equation}
\begin{split}
 \calW^\mu (x,p)
 & \equiv -\frac{1}{\hbar(p\cdot \nu)}\e^{\m\n\r\s} p_\nu M_{\rho\sigma} \,,
\end{split}
\end{equation}
where we define $M_{\rho\sigma}\equiv\nu^\lambda M_{\lambda,\rho\sigma}$ with $\nu^\mu$ being a unit time-like vector, and the factor $ 1/{p\cdot \nu} $ is introduced for normalization.
One can readily check that the above definition of $\calW^\mu$ excludes the orbital angular momentum part and can be reduced as
\begin{equation}
\label{eq:W-A}
\calW^\mu (x,p) = \calA^\mu (x,p) \,.
\end{equation}
Note that this relation is independent of the vector $\nu^\mu$.
The coincidence between $\calW^\mu$ and $\calA^\mu$ is expectable. As an example, for massless fermions, the magnetic-field induced spin polarization can be considered as the axial current generation, which is the chiral separation effect~\cite{Son:2004tq,Metlitski:2005pr}.
The spin polarization density defined with the PL vector is hence equivalent to the axial current:
\begin{equation}
 \label{eq:msW}
 \ms{W}^{\m}(x)
 \equiv\int_p\calW^\m(x,p)
 = \int_p\calA^\mu(x,p)
 = J_5^\mu(x) \,.
\end{equation}

\section{Physical Interpretation}\label{sec:phys}
In this section, we discuss the physical meanings of Eqs.~\eqref{eq:VWI}-\eqref{eq:GordonA_mag}.
To show this, we perform integration over momentum space, which results in much simpler expressions;
most of the total-derivative terms vanish as surface integrals [an exception in Eq.~\eqref{eq:AWI} is discussed later].
For simplicity, we hereafter only focus on $O(\hbar)$ terms, so that the Riemann curvature is neglected in Eqs.~\eqref{eq:VWI}-\eqref{eq:GordonA_mag} and Eq.~\eqref{ptmd} is reduced to $\Pi_\mu = p_\mu$ and $\Delta_\mu = D_\mu - F_{\mu\lambda}\partial_p^\lambda$.

First, we demonstrate that Eqs.~\eqref{eq:VWI} and~\eqref{eq:Lorentz} leads to fundamental Ward identities.
After integrating over momentum space, Eq.~\eqref{eq:VWI} gives the vector current conservation law:
\begin{equation}
\na_\m J^\m=0 \,.
\end{equation}
It is obvious from this that Eq.~\eqref{eq:VWI} is the kinetic equation of the vector charged particle.
Integrating Eq.~\eqref{eq:VWI} after multiplying by $p^\n$, we obtain the energy-momentum conservation law in the presence of the external field:
\begin{equation}
\label{emtc}
\na_\m \lb T^{\m\n} + T_{\rm ext}^{\m\n} \rb = 0 \,,
\end{equation}
where $T_{\rm ext}^{\m\n}=-F^{\m\l}F^\n_{\;\;\l} + \frac{1}{4} g^{\m\n} F^{\r\s}F_{\r\s}$ is the energy-momentum tensor of an electromagnetic field.
Here we have used Maxwell's equation $\na_\m F^{\m\n}=J^\n$ and the Bianchi identity $\na_\m \tilde{F}^{\m\n}=0$ with $\tilde{F}^{\m\n}=\frac{1}{2}\e^{\m\n\r\s}F_{\r\s}$.
On integrating Eq.~\eqref{eq:Lorentz} over momentum, we obtain
\begin{equation}
\label{samc}
\na_\m S^{\m,\r\s}= T^{\s\r}-T^{\r\s} \,.
\end{equation}
This, combined with Eq.~\eqref{emtc}, gives the conservation law of the canonical total angular momentum:
\begin{equation}
\label{amtfs}
\na_\m \lb M^{\m,\r\s} + M^{\m,\r\s}_{\rm ext} \rb = 0
\end{equation}
with $M^{\lambda,\mu\nu}_{\rm ext} = x^\mu T^{\lambda\nu}_{\rm ext} - x^\n T^{\lambda\mu}_{\rm ext}$ being the angular momentum of electromagnetic field.
This reflects the absence of the Lorentz anomaly~\cite{PhysRevLett.53.21}.

Next, we consider Eqs.~\eqref{eq:pdotA}-\eqref{eq:GordonA_mag}, the physical contents of which are less transparent than those of Eqs.~\eqref{eq:VWI} and~\eqref{eq:Lorentz}.
Equation~\eqref{eq:pdotA} involves only $\mc{A}^\m$; thus, it is a subsidiary condition for $\mc{A}^\m$. Up to $O(\hbar)$, it reduces to
\begin{equation}
p_\m \mc{A}^\m =0 \,.
\end{equation}
Based on the identification~\eqref{eq:W-A}, $\calW^\mu = \calA^\mu$, the above equation implies the following facts: spin must be either perpendicular to the momentum (i.e., for massive fermions) or parallel to the momentum (i.e., for massless fermions so that $p^2=0$ on-shell classically).
In Section~\ref{sec:qkt} we discuss the details with quantum corrections.
The electromagnetic dipole moment is derived from Eq.~\eqref{eq:aLorentz}:
\begin{equation}
\label{emdipole}
m\int_p \mc{S}_{\m\n}=-\int_p\e_{\m\n\r\s}p^\r\mc{A}^\s+\hbar\na_{[\m} J_{\n]} \,,
\end{equation}
where the first (second) term on the right-hand side represents the spin (orbital) contributions.
Equations~\eqref{eq:GordonV} and \eqref{eq:GordonA} are Gordon decompositions for the vector and axial currents.
Upon integration over momentum, they separate the convection and the gradient currents:
\ba
\label{vcfm}
 \dis
mJ^\m
&=&\int_p p^\m \mc{F}+\frac{\hbar}{2}\na_\n \int_p \mc{S}^{\m\n} \,, \\
\label{spmagmc}
 \dis
m J_5^{\m} &=& - \int_p p_\n \tilde{\mc{S}}^{\m\n}+\frac{\hbar}{2}\na^\m \int_p \mc{P} \,.
\ea
We note that the second term in Eq.~\eqref{vcfm} is the covariant form of the well-known magnetization current.
Similarly, Eqs.~\eqref{eq:GordonV_mag} and~\eqref{eq:GordonA_mag} give
\ba
\label{dualforce}
 \dis
0&=&\int_p p^\m \mc{P}+\frac{\hbar}{2} \na_\n \int_p \tilde{\mc{S}}^{\m\n} \,,\\
 \dis
\label{edforce}
 0&=&-\int_p p^\n \mc{S}_{\m\n}+ \frac{\hbar}{2} \na^\m \int_p \mc{F} \,,
\ea
where the right-hand sides are dual to those of Eqs.~\eqref{vcfm} and~\eqref{spmagmc}.
We note that $\mc{P}$ is regarded as the source of spin [see Eq.~\eqref{eq:AWI}].
Thus, Eqs.~\eqref{dualforce} and~\eqref{edforce} imply that
there do  not exist vector and axial currents carrying `magnetic charges' in Dirac theory.

Finally, we consider quantum anomalies related to Eqs.~\eqref{eq:AWI} and~\eqref{eq:dilatation}.
The momentum integral of Eq.~\eqref{eq:AWI} generates a nonvanishing surface term.
After a technical evaluation of such a term, we derive the anomalous axial Ward identity:
\ba
\na_\m J_{5}^\m
= \ms{A} -\frac{2m}{\hbar}\int_p\mc{P},
\label{eq:anomalousWI}
\ea
where $\ms{A}$ is the chiral anomaly originating from the surface integral (see Appendix~\ref{app:chrialanomaly}).
Similarly, from Eq.~\eqref{eq:dilatation}, we obtain
\begin{equation}
\label{eq:traceWI}
{T^\mu}_\mu = m \int_p \mc{F} \,,
\end{equation}
which represents the Ward identity in terms of the dilatation.
We emphasize that up to $O(\hbar)$, no surface integral contributes to Eq.~\eqref{eq:traceWI}.
As a result, the trace anomaly does not emerge here, while the chiral anomaly does as given in Eq.~\eqref{eq:anomalousWI}.
Indeed, one can confirm from dimensional analysis that the trace anomaly is $O(\hbar)$ higher than the chiral anomaly%
~\footnote{%
In this paper we neglect the $\hbar^{-3}$ in the volume element of the momentum phase space integral.
Counting such an additional power of $\hbar$, one can write the well-known anomalous Ward identities for massless fermions in Minkowski spacetime: $\partial_\mu J_5^\mu = \hbar^{-2} \frac{e^2}{16\pi^2}\epsilon^{\mu\nu\alpha\beta} F_{\mu\nu} F_{\alpha\beta}$ and ${T^\mu}_\mu = \hbar^{-1}\frac{\beta}{2e} F^{\mu\nu}F_{\mu\nu}$ with $\beta$ being the $\beta$-function (we take $c=1$ but recover $e$ explicitly).%
}.
For the same reason, the chiral anomaly in Eq.~\eqref{eq:anomalousWI} is not involved in the gravitational contribution, which is $O(\hbar^2)$ higher than the electromagnetic one~\cite{Liu:2018xip}.
In the kinetic theory involving $O(\hbar^2)$ or $O(\hbar^3)$ terms, these additional contributions should enter the right-hand sides of Eqs.~\eqref{eq:anomalousWI} and~\eqref{eq:traceWI}.
We will leave discussion of the higher order kinetic theory to a future publication.

\section{Kinetic equations at $O(\hbar)$}\label{sec:qkt}

In the kinetic theory up to $O(\hbar)$, the general solutions for $\mathcal{V}_\m$ and $\mathcal{A}_\m$ take the following forms (see Appendix~\ref{app:VandA}):
\begin{eqnarray}
\label{eq:generalV}
&&\dis
\mc{V}^\m
= 4\p
	\biggl[
		\d(\xi)
		\biggl(
			p^\m f
			+ \frac{\hbar}{2p\cdot n}\epsilon^{\m\n\r\s} n_\n \D_\r \bar{\calA}_{\s}
		\biggr)
		\non
&&\qquad\qquad\ \dis
	    +\d'(\xi)
	     \hbar \tilde{F}^{\m\n}
		 \biggl(
		 	 \bar{\calA}_{\n}
		 	 -\frac{p\cdot\bar{\calA}}{p\cdot n} n_\n
		 \biggr)
	\biggr]\,,\\
\label{eq:generalA}
&&\dis
\mc{A}^\m
= 4\p
	\biggl[
		 \delta(\xi) \bar{\calA}^{\m}
	 	 + \delta'(\xi)\hbar \tilde{F}^{\m\n}p_\n f
	 \biggr],\ \
\end{eqnarray}
where we utilize $x\delta'(x) = -\delta(x)$, and denote $\D_\m= D_\m -F_{\m\l}\partial_p^\lambda$ and
\begin{equation}
 \xi = p^2 - m^2 \,.
\end{equation}
The scalar function $f=f(x,p)$ is to be considered as the distribution function of vector charge, and $n^\m$ is a unit time-like vector that satisfies $p\cdot n\neq0$.
According to Eq.~\eqref{eq:pdotA}, the vector $\bar{\calA}^{\m}$ must satisfy the condition
\begin{equation}
\label{eqfora}
p_\m \bar{\calA}^{\m} \d(\xi)=0 \,.
\end{equation}
Here $\bar{\calA}^{\m}$ is not necessarily perpendicular to the momentum due to the presence of the delta function.
To proceed, we decompose $\bar{\calA}^{\m}$ as
\begin{equation}
\label{paraandperofa}
\bar{\calA}^\m=p_\m f_5 + \bar{\calA}_{\perp}^\m \,,
\end{equation}
where $\bar{\calA}_{\perp}^\m$ is perpendicular to the momentum: $p\cdot\bar{\calA}_\perp = 0$.

\subsection{Massless case}
In the massless limit, plugging Eqs.~\eqref{eq:generalV} and~\eqref{eq:generalA} into Eq.~\eqref{eq:aLorentz}, we identify
\begin{equation}
\label{eq:Sigman}
 \bar{\calA}_{\perp}^\m=\hbar\S_n^{\m\n}\D_\n f \,,
 \quad
 \S_n^{\m\n}\equiv\frac{\e^{\m\n\r\s}p_\r n_\s}{2p\cdot n} \,.
\end{equation}
Then, the solutions~\eqref{eq:generalV} and~\eqref{eq:generalA} are reduced to
\begin{equation}
\begin{split}
\label{eq:masslessVA}
(\mc{V},\mc{A})^\m
 & = 4\p
 \biggl[
	\d(p^2)
	\Bigl\{
		p^\m \lb f,f_5 \rb
		+\hbar \S_n^{\m\n}\D_\n \lb f_5,f \rb
	\Bigr\}  \\
 &\quad\qquad
	+ \hbar \d'(p^2) \tilde{F}^{\m\n}p_\n \lb f_5,f \rb
 \biggr] \,.
\end{split}
\end{equation}
This indicates that $f_5$ plays the role of the axial charge distribution function.
The second term in the above equation is called the side-jump term (or the magnetization current), and $\S_n^{\m\n}$ is known as the spin tensor at the spin-defining vector $n^\mu$~\cite{Chen:2014cla,Chen:2015gta}; e.g., $\Sigma^{ij}_n = \epsilon^{ijk0}p_k/2p_0$ in the rest frame $n^\mu = (1,\bzero)$.
Additionally, it is important to note from the above $\calA^\mu$ that $\Sigma_n^{\mu\nu}$ is connected to the canonical spin current~\eqref{eq:Sc} through
\begin{equation}
\label{spininckt}
 S^{\l,\m\n}n_\l
 =\hbar \int_p 4\p \d(p^2)\, p\cdot n\, f_5 \S_n^{\m\n}\,.
\end{equation}
This relation more transparently accounts for why $\Sigma^{\mu\nu}_n$ characterizes the particle spin, and  $n^\m$ represents the frame of the spin%
~\footnote{%
Strictly speaking, $n^\m$ in Eq.~\eqref{spininckt} is in a subset of the $n^\m$'s allowed to enter $\S_n^{\m\n}$.
The latter is defined in phase space while the former depends on $x$ only.%
}.

Note that Eq.~\eqref{eq:masslessVA} correctly reproduces the solution in the CKT, with the replacement as $\calR^\mu/\calL^\mu = \frac{1}{2}(\calV\pm\calA)^\mu$ and $f_{R/L}=\frac{1}{2}(f\pm f_5)$~\cite{Liu:2018xip}.
Accordingly, the chiral kinetic equations are also obtained from Eqs.~\eqref{eq:VWI} and \eqref{eq:AWI} with the above solutions~\eqref{eq:masslessVA}:
\begin{equation}
\begin{split}
\label{keml}
0 & = \d(p^2\mp\hbar F_{\a\b}\S_n^{\a\b})
	\biggl[
		p_\m \D^\m  f_{R/L} \\
  &	\quad
		\pm\frac{\hbar}{p\cdot n}\tilde{F}_{\m\n} n^\m \D^\n f_{R/L}
  		\pm \hbar \D^\m \lb \S^n_{\m\n}\D^\n f_{R/L} \rb
  	\biggr] \,.
\end{split}
\end{equation}
More discussions about the CKT can be found, e.g., in Refs.~\cite{Hidaka:2016yjf,Huang:2018wdl,Liu:2018xip}.

Now, we re-consider the chiral anomaly in the CKT.
Using the $O(\hbar)$ solution $\mc{A}^\m$ in Eq.\eqref{eq:masslessVA}, we derive the anomalous Ward identity
\begin{equation}
\begin{split}
\label{eq:msA}
 &\qquad\quad	\na_\m J^\m_{5} = \ms{A} \,,\\
 \ms{A}
 & = -\frac{\hbar}{8} F^{\mu\nu} \tilde{F}_{\mu\nu}
 		\int_\bp f(\bp) \partial_p^i \frac{p_i}{|\bp|^3} \\
 & = -\frac{\hbar}{16\pi^2} F^{\mu\nu} \tilde{F}_{\mu\nu} f(\bzero) \,,
\end{split}
\end{equation}
with $\int_\bp \equiv\int \frac{d^3 p}{(2\p)^3}$.
This reproduces the usual chiral anomaly when $f$ is (the twice of) the Fermi-Dirac distribution (see details in Appendix~\ref{app:chrialanomaly}).
The important fact is that $\ms{A}$ receives the contribution only from the singular term at $p^2=0$,  which generates the Berry monopole $\partial_p^i \frac{p_i}{|\bp|^3} = 4\pi\delta^3(\bp)$.
The above covariant expression hence manifests that the chiral anomaly is a topological nature of massless fermions in an electromagnetic field~\cite{Son:2012wh,Stephanov:2012ki}.

\subsection{Massive case}\label{subsec:mkt}
We now focus on the massive case, in which we can perform two reductions for the solutions~\eqref{eq:generalV} and~\eqref{eq:generalA}.
First, Eq.~\eqref{eqfora} for $m\neq 0$ implies
\begin{equation}
\label{eq:f5vanish}
 f_5\d(\xi)=0 \,,
\end{equation}
with which one can remove the parallel part in $\bar{\mc{A}}^\m$ from the solutions~\eqref{eq:generalV} and~\eqref{eq:generalA}.
Second, the frame vector $n^\mu$ in Eqs.~\eqref{eq:generalV} and~\eqref{eq:generalA} can be absorbed into the distribution function $f$, redefined as (see Appendix~\ref{app:removeframe}):
\ba
\label{redefmsdf}
 f \rightarrow
 f + \frac{\hbar}{m^2}\frac{\epsilon_{\m\n\r\s} p^\m n^\n}{2 p\cdot n}
 	\D^\r \bar{\calA}_{\perp}^\sigma \,.
\ea
We emphasize that this redefinition is equivalent to identifying the frame $n^\m$ as the particle's rest frame $n_\text{rest}^\m=p^\m/m$.
The frame vector $n^\m$ can be removed because, in the massive case, there is a special choice of $n^\mu$, i.e., the rest frame $n_\text{rest}^\m$.
Thus, we can always redefine the scalar distribution function $f$ from $n^\m$ to $n_\text{rest}^\m$ through a local Lorentz boost.
This procedure does not work for massless fermions due to the lack of such a special frame, making it inevitable to introduce $n^\m$.

Because of the constraint $p\cdot\bar{\calA}_\perp = 0$, there are three degrees of freedom in $\bar{\calA}_\perp^\mu$.
One of them is interpreted as the axial charge distribution, which specifies the norm of $\bar{\calA}_\perp^\mu$.
Because the axial current density is identified as the particle spin, as shown in Eq.~\eqref{eq:W-A},
the other two correspond to the parameters of the spin direction.
In the massive case, we hence parametrize $\bar\calA^\mu_\perp$ as
\begin{equation}
\bar{\calA}_{\perp}^{\m} \equiv m\h^\m f_A \,,
\end{equation}
where $f_A$ is the axial distribution function and $\theta^\mu$ is a unit vector (with two degrees of freedom).
Note that $\theta^\mu$ is normalized with the space-like condition $\h^\m\h_\m=-1$ and $p_\m\h^\m=0$.
In addition, it is useful for later discussion to introduce the following tensor:
\begin{equation}
\label{spindef}
\S_S^{\m\n}\equiv\frac{1}{2m}\e^{\m\n\r\s}\h_\r p_\s \,,
\end{equation}
which may be regarded as the spin tensor of massive fermions, as $\Sigma_n^{\mu\nu}$ is for massless fermions~\eqref{eq:Sigman}.
Indeed, it is readily checked that $\Sigma^{ij}_S = \epsilon^{ijk0}\theta_k/2$ for the rest particle with $p_\mu = (m,\bzero)$.

Collecting the discussions above, we present the solutions of the Clifford coefficients $\mc{F}$, $\mc{P}$, $\mc{V}_\m$, $\mc{A}_\m$ and $\mc{S}_{\m\n}$, as follows:
\ba
\label{solutionf}
\dis
\mc{F}
 &=& 4\p
 	\biggl[
		 \d(\xi) m f
		 - \d'(\xi)\hbar m F^{\m\n} \S^S_{\m\n} f_A
	\biggr], \\
\label{solutionp}
\dis
\mc{P}
 &=& -2\pi\hbar  \D_\m\ls\h^\m f_A \d(\xi)\rs, \\
\label{solutionmsv}
\dis
\mc{V}^\m
&=& 4\p
	\biggl[
		\d(\xi)
		\biggl(
			p^\m f
			+\hbar\frac{\epsilon^{\m\n\r\s}}{2m}p_\n \D_\r \lb \h_\s f_A\rb
		\biggr) \non
&&\dis\qquad
		+ \d'(\xi)\hbar\,m\tilde{F}^{\m\n} \h_\n f_A
	\biggr], \\
\label{solutionmsa}
\dis
\mc{A}^\m
&=& 4\p
	\biggl[
		\d(\xi) m\h^\m f_A
		 + \d'(\xi)\hbar \tilde{F}^{\m\n}p_\n f
	\biggr],\\
\label{solutions}
\dis
\mc{S}^{\m\n}
&=& 4\p
	\biggl[
		 \delta(\xi)
		 \biggl(
		 	2 m f_A \S_S^{\m\n}
			-\frac{\hbar}{m}p^{[\m}\D^{\n]}f
		 \biggr) \non
&& \dis \qquad
			 -  \delta'(\xi) \hbar m F^{\m\n} f
	\biggr]
\ea
with $\xi = p^2 -m^2$.
In Eqs.~\eqref{solutionf}-\eqref{solutions}, there are four independent variables:
two for the distribution functions $f$ and $f_A$, and the other two for the spatial orientation of the spin vector $\h^\m$.
Therefore, the covariant spin kinetic theory up to $O(\hbar)$ is described by the following four independent evolution equations:
\begin{equation}
\begin{split}
\label{keosspud}
0&=\d(\xi\mp \hbar\S_S^{\a\b} F_{\a\b}) \\
 &\ \times
\biggl[
	\biggl(
		p\cdot \D
		\pm\frac{\hbar}{2}\S_S^{\m\n}
		\lb
			\na_\r F_{\m\n}
			-p_\l{R^\l}_{\r\m\n}
		\rb \pt^\r_p
	\biggr) f_{\pm} \\
 & \qquad
 	+\frac{\hbar}{2}(f_+ - f_-)\biggl(\lb \na_\r F_{\m\n} -p_\l{R^\l}_{\r\m\n}\rb \pt^\r_p\S_S^{\m\n}\\
 & \qquad \qquad\quad
 - \frac{1}{2m} \tilde{F}^{\n\s} \pt^p_\n \lb p\cdot \D \h_\s -F_{\s\l}\h^\l \rb \biggr)
\biggr] \,,
\end{split}
\end{equation}
\begin{equation}
\begin{split}
\label{keospin}
0 &=  \d(\xi)
	\biggl[
		f_Ap\cdot\D \h^\m -
		f_AF^{\m\n} \h_\n
		+ \h^\m  p\cdot\D f_A \\
&\qquad\quad
		-\frac{\hbar}{4m }\epsilon^{\m\n\r\a} p_\a
			\lb \na_\s F_{\n\r} -p_\l{R^\l}_{\s\n\r} \rb
			\pt^\s_p f \\
&\qquad\quad
-\frac{\hbar}{2m} \tilde{F}^{\m\n} \pt^p_\n \lb p\cdot\D f \rb
	\biggr] \,,
\end{split}
\end{equation}
with $f_{\pm}\equiv \frac{1}{2} \lb f \pm f_A \rb$.
In Appendix~\ref{app:ktms}, we present the derivation of the above kinetic equations.
With given initial conditions, Eqs.~\eqref{keosspud} and~\eqref{keospin} determine the time evolutions of $f_\pm$ and $\h^\m$ for massive fermions at the collisonless limit. The flat-spacetime counterparts of Eqs.~(\ref{keosspud}) and (\ref{keospin}) were discussed recently in Refs.~\cite{Hattori:2019ahi,Gao:2019znl,Weickgenannt:2019dks}.

We provide some comments about Eqs.~\eqref{keosspud} and \eqref{keospin}: \\
{\bf(1)} $\S_S^{\m\n}$ is related to $\h^\m$ through its definition (\ref{spindef}).
Thus in Eq.~\eqref{keosspud}, it is sufficient to keep only the $O(1)$ order contribution in $\S_S^{\m\n}$, which is always accompanied by an additional $\hbar$ factor.
\\
{\bf(2)}
The delta function in Eq.~\eqref{keosspud} shows that the onshell condition is shifted by $\mp \hbar\S_S^{\a\b} F_{\a\b}$.
This term should be regarded as the magnetization coupling, similar to $\mp \hbar\S_n^{\a\b} F_{\a\b}$ in the massless kinetic equation~\eqref{keml}.
\\
{\bf(3)}
Note that $f_+$ ($f_-$) represents the distribution for fermions that have spin parallel (antiparallel) to $\h^\m$.
Indeed, the particle number of such spin-aligned fermions can be written with the Wigner function, as follows:
\begin{equation}
\begin{split}
 N_\pm
  \equiv \int_p \tr [ \ms{P}_\pm W]
  = \int_p 4\p \d(\xi\mp \hbar\S_S^{\a\b} F_{\a\b})\,m f_{\pm} \,,
\end{split}
\end{equation}
where $\ms{P}_\pm \equiv \frac{1}{2}(1\pm\g^5\g^\m\h_\m)$ is the spin projection operator in terms of $\theta^\mu$~\cite{Kaku:1993ym}.
Moreover, this observation of $f_\pm$ is consistent with Eq.~\eqref{keosspud};
the two kinetic equations of $f_\pm$ degenerate to the same Vlasov equation $\delta(\xi) p^\mu(\partial_\mu + \Gamma_{\mu\nu}^\rho p_\rho\partial_p^\nu - F_{\mu\nu}\partial_p^\nu) f_\pm = 0$ in the classical limit, where spin-up/-down particles are indistinguishable.
\\
{\bf(4)} The third term in Eq.~\eqref{keospin} is of $O(\hbar)$ order, as we check by substituting Eq.~\eqref{keosspud}.
Therefore, in the classical limit, Eq.~\eqref{keospin} is reduced to $p\cdot\D \h^\m =F^{\m\n} \h_\n$ with the on-shell condition $p^2=m^2$.
This is the the Bargmann--Michel--Telegdi (BMT) equation, which describes the Larmor-Thomas procession of the spin~\cite{Bargmann:1959gz};
in Minkowski spacetime, the BMT equation for a rest particle under a magnetic field $\bm{B}$ is written as the well-known form of the usual Larmor procession: $m\dot{\bh}=\bm{B\times\bh}$.
\\
{\bf(5)} From Eq.~\eqref{keosspud}, we extract the following single-particle equations of motion:
\ba
\label{eomforx}
\frac{D x^\m}{D\t}&=& \frac{p^\m}{m}\,,\\
\label{eomforp}
\frac{D p^\m}{D\t}&=& F^{\m\l}\frac{p_\l}{m}\pm \frac{\hbar}{2m}\S_S^{\a\b}\lb\nabla^\m F_{\a\b}-p_\l{R^{\l\mu}}_{\a\b}\rb \,.\quad
\ea
Here $D/D\t$ is the covariant derivative in terms of $\tau$, which is the proper time along the trajectory of the particle, and the on-shell condition $\xi\mp \hbar\S_S^{\a\b} F_{\a\b}=0$ is implicitly applied.
Equation~\eqref{eomforp} is known as the first Mathisson--Papapetrou--Dixon (MPD) equation~\cite{Mathisson:1937zz,Papapetrou:1951pa,Dixon:1970zza}.
The first, second, and third term in Eq.~\eqref{eomforp} represent the Coulomb-Lorentz force, the Zeeman force, and the spin curvature coupling, respectively.
\\
{\bf(6)}
Multiplying $\e_{\a\b\w\m}p^\w$, Eq.~\eqref{keospin} becomes $p\cdot \D \S_S^{\m\n}=2F_\s^{\;\;[\m}\S_S^{\n]\s}+O(\hbar)$.
Combining this with Eq.~\eqref{eomforx}, the following equation of motion is derived:
\begin{equation}
\label{eomfors}
\frac{D\hbar\S_S^{\m\n}}{D\t}
 = 2\frac{1}{m}F_\s^{\;\;[\m}\hbar\S_S^{\n]\s}+2p^{[\m}\frac{Dx^{\n]}}{D\t} \,.
\end{equation}
This is the second MPD equation, which determines the spin motion in electromagnetic and gravitational backgrounds~\cite{Mathisson:1937zz,Papapetrou:1951pa,Dixon:1970zza}.
Note that the Tulczyjew-Dixon condition~\cite{Tulczyjew,dixon1974dynamics} is automatically satisfied: $p_\m\S_S^{\m\n}=0$.

\section{Application: spin polarization}\label{sec:sppol}

\subsection{General state}\label{sec:general}
As an application of our spin kinetic theory, we calculate the spin polarization of Dirac fermions, which is an intensively studied topic in heavy-ion collisions.
As already mentioned, an unambiguous definition of the spin polarization is the PL vector $\calW^\mu = \calA^\mu$ in Eq.~\eqref{eq:W-A} and $\ms{W}^\mu(x) = \int_p \calW^\mu = \int_p \calA^\mu$ in Eq.~\eqref{eq:msW}.
Combined with Eqs.~\eqref{eq:masslessVA} and~\eqref{solutionmsa}, this polarization vector is expressed by
\begin{equation}
\label{eq:msW_general}
\ms{W}^{\m} (x) =
\begin{cases}
\dis
 \int_p 4\pi\d(p^2)
 	\biggl[
 		p^\m f_5
 		+ \hbar \S_n^{\m\n}\D_\n f
 		- \frac{\hbar}{2} \tilde{F}^{\m\n}\pt^p_\n f
 	\biggr]\\
 \qquad\qquad\qquad\qquad\qquad\qquad\qquad \text{(massless)\,,}
 \\
\dis
 \int_p 4\pi\d(\xi)
 	\biggl[
 		m\h^\m f_A
 		- \frac{\hbar}{2} \tilde{F}^{\m\n}\pt^p_\n f
 	\biggr] \\
 \qquad\qquad\qquad\qquad\qquad\qquad\qquad \text{(massive)\,.}
\end{cases}
\end{equation}
For later use, we also define the polarization per particle in the phase space:
\begin{equation}
\label{eq:w_general}
 w^\mu (x,p) = \frac{\calW^\mu(x,p)}{4\pi f(x,p)} =  \frac{\calA^\mu(x,p)}{4\pi f(x,p)} \,.
\end{equation}
These expressions are available in nonequilibrium state.
The last terms in each case stem from the Zeeman coupling, which gives an additional $O(\hbar)$ contribution.
In addition to the magnetic field, other sources, such as the fluid vorticity (or rotation), also induces spin polarization.
In Eq.~\eqref{eq:msW_general}, such contributions are found, only after the concrete forms of the distribution functions are determined.
For this analysis, the collision terms are needed, which we will discuss in a subsequent paper.
In global equilibrium state, however, we can identify the vorticity-dependence of the distribution functions without knowing the collision terms, as shown below.

\subsection{Equilibrium state}\label{sec:eqst}
In this section, we study spin polarization in the equilibrium state.
In kinetic theory, the {\it local equilibrium state} is specified by the distribution functions that eliminate the collision kernel.
This implies that the distribution functions must depend only on the linear combination of the collisional conserved quantities: the particle number, the energy and momentum, and the angular momentum.
Therefore, we consider the following ansatz for the local equilibrium distributions, $f^{\rm LE}_{\pm}=n_F(g_{\pm})$ with $g_{\pm}=p \cdot \b + \a_\pm \pm \hbar\S_S^{\m\n}\o_{\m\n}$ for massive fermions (where we have absorbed the orbital angular momentum into a redefinition of the $\b$ field), and $f^{\rm LE}_{R/L}=n_F(g_{R/L})$ with $g_{R/ L}=p \cdot \b + \a_{R/ L} \pm \hbar\S_n^{\m\n}\o_{\m\n}$ for massless fermions. The coefficients $\b_\m, \a$'s, $\o_{\m\n}$ (called spin chemical potential) depend only on $x$, where $\b^\m$ is assumed to be time-like.
Although the actual functional form of $n_F$ is not essential, we assume it to be the Fermi-Dirac function for demonstration.

\subsubsection{Massive case}\label{sec:eqst:massive}
In the massive case, at local equilibrium, the spin polarization vectors are readily computed from Eqs.~\eqref{eq:msW_general} and~\eqref{eq:w_general}, as follows:
\begin{eqnarray}
\label{eq:wLE}
 &\dis w^\mu_\LE (x,p)
 = - \delta(\xi) m\theta^\mu(\alpha_A + \hbar\theta\cdot\Omega)\bar{n}_F
 	+ \hbar\delta'(\xi)\tilde{F}^{\mu\nu}p_\nu \,,\non \\
\label{spdmv}
 &\dis  \ms{W}_\LE^{\m}(x)
= 4\p\int_p\d(\xi)
	\Bigl[
		2m\h^\m(\a_A + \hbar\Omega\cdot\theta)
		-\hbar \tilde{F}^{\m\n}\b_\n\Bigr] n_{F}' \,, \non
\end{eqnarray}
with $n_F=n_F(p\cdot\b+\a)$, $\bar{n}_F = 1- n_F$, $\a=(\a_++\a_-)/2$, $\a_A=(\a_+-\a_-)/2$, and $\Omega^\m=\e^{\m\n\r\s}p_\n\o_{\r\s}/(2m)$.
Note that $\alpha_A$ is assumed to be of $O(\hbar)$; otherwise a finite spin polarization would be generated, even in the classical limit.

It is more important to discuss the polarization at global equilibrium.
For this purpose, we determine necessary constraints imposed by the kinetic equations~\eqref{keosspud} and~\eqref{keospin}.
Substituting $f^{\rm LE}_{\pm}$ into Eq.~\eqref{keosspud}, one can show that the following conditions
can fulfil Eq.~\eqref{keosspud} up to $O(\hbar)$ for an arbitrary spin vector~$\h^\m$:
\begin{eqnarray}
 \label{eq:eq_killing}
&\na_\m \b_\n + \na_\n \b_\m = 0,\\
 \label{eq:eq_vorticity}
&\na_{[\m} \b_{\n]} - 2\o_{\m\n} = 0,\\
 \label{eq:eq_alpha}
&\na_\m \a_{\pm} =F_{\m\n}\b^\n \,.
\end{eqnarray}
Furthermore, we verify that under the conditions~\eqref{eq:eq_killing}-\eqref{eq:eq_alpha}, the following choice of $\a_A$ and $\h^\m$ fulfills Eq.~\eqref{keospin} (see Appendix \ref{app:eqforspin}):
\begin{equation}
\begin{split}
\label{eqspinvec}
\a_A=0\,, \quad
\h^\m=-\frac{1}{2m\G}\e^{\m\n\r\s}p_\n\nabla_{\r}\b_{\s},
\end{split}
\end{equation}
where $\G=(\frac{1}{2}\na_{[\m}\b_{\n]}\L^{\m\r}\L^{\n\s}\na_{[\r}\b_{\s]})^{1/2}$ with $\L^{\m\n}=g^{\m\n}-p^\m p^\n/m^2$.
We call the state specified by the conditions~\eqref{eq:eq_killing}-~\eqref{eqspinvec} the {\it global equilibrium state} and denote $f^\GE$ as the corresponding distribution function.
At global equilibrium, the thermal vorticity $\na_{[\m} \b_{\n]}$ determines both the spin chemical potential $\o_{\m\n}$ and the spin vector $\h^\m$. We emphasize that finite Riemann curvature or an external electromagnetic field is necessary to derive Eq.~\eqref{eq:eq_vorticity}. Without the external electromagnetic field and the curved background geometry, the spin degree of freedom is inactive in the collisionless kinetic theory and we cannot link $\o_{\m\n}$ to $\na_{[\m} \b_{\n]}$.
Additionally, in Appendix~\ref{app:denop}, we re-derive the conditions~\eqref{eq:eq_killing}-\eqref{eq:eq_alpha} for massive fermions [and \eqref{eq:eqml_killing}-\eqref{eq:eqml_alpha} for massless fermions] based on the density operator.

At global equilibrium, the spin polarization vectors read
\begin{eqnarray}
\label{eq:wGEm}
& \dis
 w^\mu_\GE(x,p)
 = \hbar\frac{\delta(\xi) }{2} \tilde\omega^{\mu\nu} p_\nu \bar{n}_F
 	+ \hbar\delta'(\xi) \tilde{F}^{\mu\nu} p_\nu \,, \\
\label{spdmv2}
& \dis
\ms{W}_\GE^{\m}(x)
= 4\p\hbar \int_p \d(\xi) \ls - \tilde{\omega}^{\mu\nu}p_\nu - \tilde{F}^{\m\n}\b_\n\rs n_{F}'
\end{eqnarray}
with $\tilde{\omega}^{\mu\nu} = \frac{1}{2}\epsilon^{\mu\nu\rho\sigma}\omega_{\rho\sigma}$.
In addition to $\calW^\mu$ and $w^\mu$, at global equilibrium, it is also practically useful to compute the space-integrated polarization.
Suppose that the fermions are frozen out on a space-like hypersurface $\S^\m(x)$.
The average spin polarization per particle after freeze-out is given by%
~\footnote{Here we pick up the particle-branch contribution.
The anti-particle-branch contribution are similarly obtained by replacing $\int_0^\infty d(p\cdot u)$ by $\int_{-\infty}^0 d(p\cdot u)$.}
\begin{equation}
\begin{split}
\label{freezeout}
\bar{\calW}_\GE^{\m}(p)
&\equiv \frac{\int d\S^\l p_\l \int_0^\infty d(p\cdot u)\calW^\m_\GE(x,p)}{4\p\int d\S^\l p_\l f_\GE(x,p)}\\
&=
\frac{\int d\S^\l p_\l
	\frac{\hbar}{4E_p}\ls
		-\tilde{\omega}^{\mu\nu}p_\nu
		- \tilde{F}^{\m\n}\b_\n
		\rs n_{F}'}{\int d\S^\l p_\l n_F} \,.
\end{split}
\end{equation}
If we set $F_{\mu\nu} = 0$, the above equation is consistent with the result derived in Refs.~\cite{Becattini:2013fla,Fang:2016vpj}~\footnote{Note that the spin polarization is defined as $\bar{\calW}_\GE^{\m}(p)/s$ with $s=1/2$ the spin quantum number therein.}, which has been widely used for the calculation of the hadron spin polarization.
In the above $u^\m=T\b^\m$ is the fluid velocity~\footnote{Note that $\b^\m$ is a Killing vector owing to Eq.~\eqref{eq:eq_killing}.} and the momentum in the second line is on-shell; in Minkowski spacetime and in the local rest frame of the fluid, $p^\m=(E_p=\sqrt{\bm{p}^2+m^2},\bm{p})$ where $\bm{p}$ is the three momentum.

\subsubsection{Massless case}\label{sec:eqst:massless}
In the same manner, Eq.~\eqref{keml} with $f^{\rm LE}_{R/L}$ yields the following global equilibrium conditions~\cite{Liu:2018xip}:
\begin{eqnarray}
\label{eq:eqml_killing}
 &\na_\m \b_\n + \na_\n \b_\m = \f (x) g_{\m\n},\\
\label{eq:eqml_vorticity}
 &\na_{[\m} \b_{\n]} - 2\o_{\m\n} = 0,\\
\label{eq:eqml_alpha}
 &\na_\m \a_{R/L}=F_{\m\n}\b^\n \,.
\end{eqnarray}
Unlike the massive case, the first conditions has an arbitrary function $\f (x)$, which arises as a result of the conformal invariance in the massless case; thus, $\b^\m$ is a conformal Killing vector.
At global equilibrium, analogously to Eqs.~\eqref{eq:wGEm}-\eqref{freezeout}, we calculate
\begin{equation}
 w^\mu_\GE(x,p)
 = \frac{\hbar\delta(p^2)}{2}\bigl(- 2p^\m \a_5/\hbar + \tilde{\omega}^{\mu\nu}p_\nu \bigr)\bar{n}_F
   + \hbar\delta'(p^2) \tilde{F}^{\mu\nu} p_\nu \,,
\end{equation}
\begin{equation}
\begin{split}
\ms{W}_\GE^\m (x)
 = 4\p\hbar \int_p\d(p^2) \Big[ 2p^\m \a_5/\hbar - \tilde{\omega}^{\mu\nu}p_\nu  - \tilde{F}^{\m\n}\b_\n  \Big]n_F'\,,
\end{split}
\end{equation}
\begin{equation}
\begin{split}
& \bar{\calW}_\GE^{\m}(p)
 = \frac{\int d\S^\l p_\l \frac{\hbar}{4E_p}
		\Big[
			2p^\m\a_5/\hbar
			- \tilde{\omega}^{\mu\nu}p_\nu
			- \tilde{F}^{\m\n}\b_\n
		\Big] n_{F}'}{\int d\S^\l p_\l n_F}
\end{split}
\end{equation}
with $\alpha_5 = (\alpha_R - \alpha_L)/2$, which is of $O(\hbar)$ as well as $\alpha_A$%
~\footnote{%
More precisely, while the kinetic equation~\eqref{keosspud} requires $\a_A$ to be of $O(\hbar)$ in the global equilibrium, there is no such requirement for $\a_5$.
Nevertheless, one should assume that $\alpha_5$ is of $O(\hbar)$; otherwise a finite axial charge would be generated even in the classical limit.%
}.
In the second equation, the on-shell condition is implicitly applied and we define $E_p=u\cdot p$;
in Minkowski spacetime and the rest frame of the fluid, $E_p=|\bm{p}|$.
Note that spin polarization induced by the thermal vorticity and the electromagnetic field takes the same form for both massless and massive cases at global equilibrium, up to the difference in the on-shell conditions.
Moreover, the results are independent of the choice of the frame vector $n^\m$, as they should be.

\section{Summary and outlook}\label{sec:sum}
In this paper, we derive the collisionless covariant spin kinetic theory at $O(\hbar)$ for Dirac fermions in curved spacetime and an external electromagnetic field.
We start by deriving the dynamical equation for each Clifford component of the Winger function up to $O(\hbar^2)$.
We discuss the physical meaning of each such dynamical equation.
We then take $\mc{V}^\m$ and $\mc{A}^\m$ as independent dynamical variables and derive two evolution equations for massless fermions~\eqref{keml} and four evolution equations for massive fermions~\eqref{keosspud} and~\eqref{keospin}, respectively.
We introduce a time-like unit frame-choosing vector $n^\m$ to solve the Wigner function.
In the massless case, $n^\m$ is necessary because it represents the frame in which the spin for the massless particle is defined.
In the massive case, we show that the vector $n^\m$ can be removed by redefining the vector distribution function through a boost from the frame $n^\m$ to the rest frame of the particle.

As an application, we analyze spin polarization using the approach of the kinetic theory.
We derive the global equilibrium conditions from the kinetic equations and find that the finite Riemann curvature or an external electromagnetic field is necessary to determine the spin-thermal vorticity coupling.
We derive the expression of spin polarization induced by the electromagnetic field and the thermal vorticity at global equilibrium, which is consistent with the results in previous literature. We also derive expressions for spin polarization at local equilibrium and out of equilibrium. They may be used to study the local $\L$ polarization puzzle found in heavy-ion collisions, which cannot be understood in the calculations based on global equilibrium assumption.

We expect the spin kinetic theory to be useful for the study of both the electromagnetic plasma and quark-gluon plasma in heavy-ion collisions.
Furthermore, formulating the kinetic theory in curved spacetime may find fundamental applications in astrophysics and condensed matter physics.
For example, our present theory may be used to study the deformed crystal or a material with a temperature gradient, which is described as an electron system in a fictitious gravity~\cite{PhysRev.135.A1505,Dong:2018zoq}. Potentially, we could study the mass correction to the chiral magnetic effect and the generation and transport of spin currents in such systems.
Numerical works to solve the kinetic theory and to simulate the evolution of spin polarization in heavy-ion collisions are also important tasks.
Once the collision term is included, it would be interesting to derive the covariant spin hydrodynamics~\cite{Florkowski:2017ruc,Florkowski:2018fap,Hattori:2019lfp} from the covariant spin kinetic theory.

\begin{acknowledgments}
We are grateful to Francesco~Becattini, Gaoqing~Cao, Ren-Hong~Fang, Lan-Lan~Gao, Xingyu~Guo, Koichi~Hattori, Yoshimasa~Hidaka, An-Ping~Huang, Jin-Feng~Liao, Xin-Li~Sheng, Qun~Wang, Xiao-Liang~Xia, Di-Lun~Yang, and Pengfei~Zhuang for useful discussions.
We also thank the Yukawa Institute for Theoretical Physics in Kyoto University, where this work was developed during the cuorse `Quantum kinetic theories in magnetic and vortical fields'.
X.-G.~H. was supported by NSFC through Grants No.~11535012 and No.~11675041, and K.~M. was supported by the China Postdoctoral Science Foundation through Grant No.~2017M621345.
\end{acknowledgments}

\begin{appendix}

\section{Derivation of chiral anomaly}\label{app:chrialanomaly}
In this Appendix, we derive the chiral electromagnetic anomaly from the solutions of the Wigner function.
We consider the massless case for demonstration.
Plugging $\calA^\mu$ in Eq.~\eqref{eq:masslessVA} into the kinetic equation~\eqref{eq:AWI}, and integrating it, we get $\na_\m J^\m_{5} = \ms{A}$ with
\begin{equation}
\label{eq:msA}
\begin{split}
 \ms{A} & = \int_p 4\pi
 	F^{\m\lambda}\pt^p_\lambda
 	 \biggl[
 	 	\hbar \d'(p^2) \tilde{F}_{\m\nu}p^\nu f \\
 & \quad
	    +\d(p^2)
	    	\biggl(
  				p_\m f_5
	    		+ \hbar\frac{\e_{\m\n\r\s}}{2p\cdot n}n^\n p^\s \Delta^\rho f
	    	\biggr)
  	\biggr] \\
 & = \frac{\hbar}{4} F^{\m\n}\tilde{F}_{\m\n}
 	\int_p  4\pi \,\d'(p^2)  p^\lambda \pt^p_\lambda f \,,
\end{split}
\end{equation}
where we employ the Schouten identity and $x\delta''(x) = - 2\delta'(x)$ and drop the surface terms without the singularity at $p^2=0$. We have chosen the local Lorentz coordinate to perform the computation as $\ms{A}$ is a scalar.
The roots of $p^2 = 0$ are $p_0 = \pm |\bp|$, with which the delta function is reduced to
\begin{equation}
 \delta(p^2)
 = \frac{1}{2|\bp|}
 	\Bigl[
 		 \delta(p_0 - |\bp|)
 		 + \delta(p_0 + |\bp|)
 	\Bigr] \,.
\end{equation}
Furthermore, when we carry out the $p_0$-integration, we need the replacement of the momentum derivatives, as follows:
\begin{equation}
\begin{split}
  \partial_p^i f(\pm|\bp|,p_i)
  & = \biggl(
  		\partial_p^i
 		+ \frac{\partial p_0}{\partial p_i}\partial_p^0
 	  \biggr) f(p_0,p_i) \Big |_{p_0 = \pm|\bp|} \\
  & = \biggl(
  		\partial_p^i
 		- \frac{p^i}{p^0}\partial_p^0
 	  \biggr) f(p_0,p_i) \Big |_{p_0 = \pm|\bp|} \\
  & \equiv \tilde{\partial_p^i}  f(p_0,p_i) \Big |_{p_0 = \pm|\bp|} \,.
\end{split}
\end{equation}
Subsequently, the integral in Eq.~\eqref{eq:msA} is cast into
\begin{equation}
\begin{split}
 & \int_p 4\pi\delta'(p^2) p^\lambda \partial^p_\lambda f \\
 & = \int_p  4\pi \frac{1}{2}\Bigl[\partial^\lambda_p \d(p^2)\Bigr] \pt^p_\lambda f
  = - \frac{1}{2} \int_p 4\pi \d(p^2) \partial^\lambda_p \pt^p_\lambda f  \\
 & = - \frac{1}{2} \int_p 4\pi\delta(p^2)
 	 \biggl[
 	 	 \tilde{\partial}_p^i \tilde{\partial}^p_i
 	 	 + \frac{2}{p_0} \partial^p_0
 	 	 +\frac{2}{p_0} p^i\tilde{\partial^p_i} \partial^p_0
 	 \biggr] f \,.
\end{split}
\end{equation}
In the last line, the second and third terms in the integrand cancel out:
performing the integration by parts, we rewrite the third term as
\begin{equation}
\begin{split}
 \int_p \delta(p^2) \frac{p^i}{p_0}  \tilde{\partial}_i^p \partial^p_0 f
 & = \int_p \delta(p^2)
 	\biggl[
 		\frac{p^i}{p_0} \partial^p_i
 		- \frac{p^i p_i}{p_0 p^0} \partial_p^0
 	\biggr] \partial^p_0 f \\
 & = - \int_p
 		\delta(p^2) \frac{1}{p_0} \partial^p_0  f \,.
\end{split}
\end{equation}
Finally, $\ms{A}$ in Eq.~\eqref{eq:msA} is calculated as
\begin{equation}
\begin{split}
\label{eq:anomaly}
 \ms{A}
 & = - \frac{\hbar}{8} F^{\m\n}\tilde{F}_{\m\n}
 		\int_\bp \frac{1}{|\bp|} \partial_p^i \partial^p_i f \\
 & = -\frac{\hbar}{16\pi^2} F^{\mu\nu} \tilde{F}_{\mu\nu} f(\bp = {\boldsymbol 0}) \,,
\end{split}
\end{equation}
where we utilize
\begin{equation}
 \partial_p^i \frac{p_i}{|\bp|^3} = 4\pi\delta^3 (\bp) \,.
\end{equation}
The usual chiral anomaly relation is recovered; hence, we take $f(\bp=\bzero)=2$\,, where the factor $2$ accounts for the spin degeneracy of Dirac fermions.

\section{General solutions at $O(\hbar)$}\label{app:VandA}
We parametrize the perturbative solutions as
\begin{equation}
 \calV^\mu = \calV^\mu_\zero +\hbar\calV^\mu_\one + O(\hbar^2)\,,\quad
 \calA^\mu = \calA^\mu_\zero +\hbar\calA^\mu_\one + O(\hbar^2) \,.
\end{equation}
According to Eqs.~\eqref{eq:dilatation}, \eqref{eq:pdotA} and \eqref{eq:GordonV}, the general solutions in the classical limit are given by
\ba
\label{msvc}
\mc{V}_{\zero}^\m &=& 4\p p^\m f^{\zero} \d(\xi) \,,
\\
\label{msac}
\mc{A}_{\zero}^\m &=& 4\p \bar{\calA}_{\zero}^\m \d(\xi)
\ea
with $\xi = p^2 - m^2$.
Here $f^{\zero}=f^{\zero}(x,p)$ is the classical vector charge distribution function and the vector $\bar{\calA}_{\zero}^{\m}$ satisfies the condition $ p_\m \bar{\calA}_{\zero}^{\m} \d(\xi)=0$.
Substituting Eqs.~\eqref{msvc} and~\eqref{msac} into Eqs.~\eqref{eq:Lorentz}-\eqref{eq:GordonA_mag}, we obtain the solutions at $O(\hbar)$:
\begin{equation}
\begin{split}
\mc{V}_\one^\m
 &= 4\p \bigg[ \lb p^\m f^\one +\frac{1}{2p\cdot n}\epsilon^{\m\n\r\s} n_\n \D_\r \bar{\calA}^{\zero}_\s \rb \d(\xi) \\
 & \qquad\qquad
 +\tilde{F}^{\m\n} \lb \bar{\calA}^{\zero}_\n-\frac{p\cdot\bar{\calA}^{\zero}}{p\cdot n} n_\n  \rb \d'(\xi)\bigg] \,,\\
\mc{A}_\one^\m
	&= 4\p \Big[ \bar{\calA}_\one^{\m} \d(\xi)
	+ \tilde{F}^{\m\n}p_\n f^{\zero} \d'(\xi)\Big]\,,\\
\end{split}
\end{equation}
where $f^\one=f^\one(x,p)$ is the first order quantum correction to the vector distribution function and $\bar{\calA}_\one^{\m}$ satisfies the same condition as $\bar{\calA}_{\zero}^{\m}$ : $ p_\m \bar{\calA}_\one^{\m} \d(\xi)=0$.
Defining $f\equiv f^{\zero}+\hbar f^\one$ and $\bar{\calA}^{\m}\equiv\bar{\calA}_{\zero}^{\m}+\hbar\bar{\calA}_\one^{\m}$, we obtain Eqs.~\eqref{eq:generalV} and~\eqref{eq:generalA}.

\section{Elimination of $n^\mu$}\label{app:removeframe}
Here, we show that, with the redefinition of the distribution function $f$ in Eq.~\eqref{redefmsdf}, we can remove the frame vector $n^\m$ from the spin kinetic theory for massive fermions. The discussion is kept at $O(\hbar)$.
Acting on Eq.~\eqref{eq:aLorentz} with $\D_\a$, we derive
\begin{equation}
\begin{split}
& p\cdot \D \mc{A}^\b + F^{\a\b}\mc{A}_\a-p^\b \D_\a\mc{A}^\a \\
&=\frac{m}{2}\e^{\a\b\r\s}\D_\a\mc{S}_{\r\s}-\frac{\hbar}{2}\e^{\a\b\r\s}\D_\a\D_\r \mc{V}_\s \,.
\end{split}
\end{equation}
Using Eqs.~\eqref{eq:AWI} and~\eqref{eq:GordonV_mag}, we obtain
\ba
p^\b \D_\a\mc{A}^\a
=\frac{m}{2}\e^{\b\a\r\s}\D_\a\mc{S}_{\r\s}.
\ea
Combining the above two equations, we find
\ba
p\cdot\D \mc{A}_\m&=&F_{\m\n}\mc{A}^\n+\frac{\hbar}{2}\e_{\m\n\r\s}\D^\n\D^\r\mc{V}^\s \,.
\label{spinevol}
\ea
Next, we substitute the redefined distribution function in Eq.~\eqref{redefmsdf} into the solution of $\mc{V}^\m$ in Eq.~\eqref{eq:generalV}, and obtain
\ba
\mc{V}^\m
&=& 4\p\d(\xi) \bigg[ p^\m \lb f - \frac{\hbar\e^{\a\b\r\s} p_\a n_\b}{2m^2 p\cdot n} \D_\r \bar{\calA}_{\perp\s} \rb \non
&& +\frac{\hbar\epsilon^{\m\n\r\s} n_\n}{2p\cdot n} \D_\r \bar{\calA}^\bot_\s \bigg]
	+4\p \d'(\xi)\hbar\tilde{F}^{\m\n} \bar{\calA}^\bot_\n \,.
\label{vsolbredf1}
\ea
To reduce the above equation, we utilize the Schouten identity:
\begin{equation}
\begin{split}
& p^\m \e^{\a\b\r\s} p_\a n_\b \D_\r \bar{\calA}_{\perp\sigma}\\
&= - \Bigl(
		p^2\e^{\b\r\s\m} n_\b \D_\r
		+ p\cdot n \e^{\r\s\m\a} p_\a \D_\r \\
&\qquad
	 	+ \e^{\s\m\a\b} p_\a n_\b p\cdot\D
	 	+ \e^{\m\a\b\r} p_\a n_\b p^\s \D_\r
	 \Bigr) \bar{\calA}_{\perp\sigma} \,,
\end{split}
\end{equation}
where the last two terms cancel, according to Eq.~\eqref{spinevol}.
We then rewrite Eq.~\eqref{vsolbredf1} as the $n^\m$ independent form:
\begin{equation}
\mc{V}^\m
 = 4\p\delta(\xi)
 \bigg[
 	p^\m  f
 	+\frac{\hbar\epsilon^{\m\n\r\s} p_\n}{2m^2} \D_\r \bar{\calA}^\bot_\s
	 -\frac{\hbar}{\xi}\tilde{F}^{\m\n} \bar{\calA}^\bot_\n
 \bigg] \,.
\end{equation}
The frame vector $n^\m$ is also removed from the solution of $\mc{A}^\m$.
Comparing the above equation with Eq.~\eqref{eq:generalV}, we find that the redefinition of $f$ is equivalent to replacing $n^\m$ with $p^\m/m$ in Eq.~\eqref{eq:generalV}.

\section{Derivation of Eqs.~\eqref{keosspud} and~\eqref{keospin}}\label{app:ktms}
Here, we derive the kinetic equations for $f_{\pm}$. Substituting Eq.~\eqref{solutionmsv} into Eq.~\eqref{eq:VWI}, one obtains the following equation:
\begin{equation}
\begin{split}
0
&=p^\m \D_\m f \d(\xi)
- \hbar\S_S^{\a\b} F_{\a\b} p^\m \D_\m f_A \d'(\xi)\\
&\quad +\frac{\hbar}{2}\lb \na_\r F_{\m\n} \pt^\r_p+\ls D_\m, D_\n \rs\rb \lb \S_S^{\m\n} f_A \rb \d(\xi)\\
&\quad
- \hbar\frac{1}{2m} f_A \tilde{F}^{\n\s} \pt^p_\n \lb p^\r \D_\r \h_\s -F_{\s\l}\h^\l \rb \d(\xi).
\end{split}
\end{equation}
In addition, contracting Eq.~\eqref{spinevol} with $\h^\m$ and inserting Eqs.~\eqref{solutionmsv} and~\eqref{solutionmsa}, we obtain
\begin{equation}
\begin{split}
0
&=p^\m \D_\m f_A \d(\xi)
- \hbar\S_S^{\a\b} F_{\a\b} p^\m \D_\m f \d'(\xi)\\
&+\frac{\hbar}{2}\S_S^{\m\n} \lb \na_\r F_{\m\n} \pt^\r_p+\ls D_\m, D_\n \rs \rb  f \d(\xi).
\end{split}
\end{equation}
The addition and subtraction of the above two equations result in Eq.~\eqref{keosspud}.
Moreover, the kinetic equation to determine $\theta^\mu$ is obtained from Eq.~\eqref{spinevol} with the solutions~\eqref{solutionmsv} and~\eqref{solutionmsa}.

\section{Global equilibrium condition from kinetic theory}\label{app:eqforspin}
In the massless case, the discussion of the global equilibrium conditions~\eqref{eq:eq_killing}-\eqref{eq:eq_alpha} was given in Ref.~\cite{Liu:2018xip}. Following a similar strategy, one can show that, for the massive case, the conditions~\eqref{eq:eq_killing}-\eqref{eq:eq_alpha} can fulfill Eq.~\eqref{keosspud} for arbitrary $\h^\m$ and for $\a_A=O(\hbar)$. Also, it is easy to see that the condition~\eqref{eqspinvec} fulfills Eq.~\eqref{keosspud} up to $O(\hbar)$.
We verify that the condition~\eqref{eqspinvec} also fulfills Eq.~\eqref{keospin} under the conditions~\eqref{eq:eq_killing}-\eqref{eq:eq_alpha}, as follows.
The leading order of $f_A^{\rm LE}$ is written as $f^\LE_A=2(\a_A+\hbar\S^S_{\a\b}\o^{\a\b}) n_F'(\b\cdot p+\a) + O(\hbar^2)$.
Using Eqs.~\eqref{eq:eq_killing}-\eqref{eqspinvec} and inserting $f^\LE$ and $f_A^\LE$, we obtain
\begin{equation}
\begin{split}
\label{spineqconapp}
 & \text{RHS of Eq.~\eqref{keospin}} \\
 & = 2 \d(\xi) n'_F
 	\biggl[
		\hbar p\cdot\D \lb \h^\m \S^S_{\a\b}\o^{\a\b} \rb
		- \lb\a_A+\hbar \S^S_{\a\b}\o^{\a\b} \rb F^{\m\n} \h_\n \\
& 	\qquad\qquad
		- \frac{\hbar}{4m}\epsilon^{\m\n\r\a} p_\a \lb \na_\s F_{\n\r}
		-p_\l{R^\l}_{\s\n\r} \rb \b^\sigma 	
		\biggr] \\
 & = - \hbar\frac{\d(\xi)}{2m} n'_F
	\Bigl[
		p\cdot\D \lb \e^{\m\n\r\s}p_\n\nabla_{\r}\b_{\s} \rb
		- F^{\m\n} \e_{\n\l\r\s}p^\l\nabla^{\r}\b^{\s} \\
& 	\qquad\qquad
		+ \epsilon^{\m\n\r\a} p_\a \lb \na_\s F_{\n\r} -p_\l{R^\l}_{\s\n\r} \rb \b^\s
	\Bigr] \,.
\end{split}
\end{equation}
In the above equation, the second equality follows from $\alpha_A = 0$ and
\begin{equation}
 \theta^\mu\Sigma^{\alpha\beta}_S\omega_{\alpha\beta}
 = \frac{1}{2} \theta^\mu \Gamma
 = -\frac{1}{4m} \epsilon^{\mu\nu\rho\sigma}p_\nu\nabla_\rho\beta_\sigma \,,
\end{equation}
which stems from Eq.~\eqref{eqspinvec}.
The above three terms in Eq.~\eqref{spineqconapp} totally vanish, as follows:
\begin{equation}
\begin{split}
 & p\cdot\D \lb \e^{\m\n\r\s}p_\n\nabla_{\r}\b_{\s} \rb  \\
 &= p_\l \lb \e^{\n\r\s\l}F^{\m}_{\;\;\n} + 2\e^{\s\l\m\n}F^{\r}_{\;\;\n} \rb \nabla_{\r}\b_{\s} \\
 & \quad +\e^{\m\n\r\s}p_\n p_\l R^\l_{\;\;\a\r\s}\b^\a  \\
 &= F^{\m}_{\;\;\n} \e^{\n\r\s\l}p_\l \nabla_{\r}\b_{\s}
 	 -p_\l \e^{\m\n\s\l} \b\cdot\na F_{\n\s} \\
 & \quad + \e^{\m\n\r\s}p_\n p_\l R^\l_{\;\;\a\r\s}\b^\a \,.
\end{split}
\end{equation}
In the above equation, we have used the Schouten identity and the equilibrium conditions~\eqref{eq:eq_killing} and ~\eqref{eq:eq_alpha} with $\nabla_{\mu} \nabla_{[\nu}\beta_{\rho]} = - \beta^\lambda R_{\lambda\mu\nu\rho}$.

\section{Global equilibrium condition from density operator}\label{app:denop}

We discuss the global equilibrium using the maximum entropy principle, following Refs.~\cite{zubarev1974nonequilibrium,Becattini:2012tc,Becattini:2019poj}.
The density operator for the local equilibrium state can be written as
\begin{equation}
\hat{\r}_\LE \equiv \frac{1}{z}
	e^{-\int d\S_\m\left(\hat{T}^{\m\n} \b_\n
		+ \hat{S}^{\m,\l\n} \o_{\l\n} +\a \hat{J}^\m\right)}\,,
\end{equation}
where $z \equiv \Tr \ls e^{-\int d\S_\m\left(\hat{T}^{\m\n} \b_\n+ \hat{S}^{\m,\l\n} \o_{\l\n}
+\a \hat{J}^\m\right)} \rs$ where $\hat{T}^{\m\n}$, $\hat{S}^{\m,\l\n}$, and $\hat{J}^\m$ are the canonical energy-momentum, spin, and charge current operators, respectively.
Here, $\S_\m$ is a space-like hypersurface, $\b^\m$, $\a$, and $\o_{\m\n}$ have the same meanings as in the main text.
The entropy is defined as
\begin{equation}
S \equiv
-\langle\ln \hat{\r}_\LE\rangle
= - \Tr \lb \hat{\r}_\LE \ln \hat{\r}_\LE \rb \,.
\end{equation}
We denote $\ln z=\int d\S_\m \f^\m$, where $\f^\m$ is (the negative of) the thermodynamic potential density current.
Then, the entropy is represented as $S=\int d\S_\m s^\m$, with
\begin{equation}
s^\m =\f^\m +  T^{\m\n} \b_\n + S^{\m,\l\n} \o_{\l\n} + \a J^\m \,.
\end{equation}
The global equilibrium condition is such that the local thermodynamic potential and entropy are maximized, so that $\na_\m \f^\m = 0$ and $\na_\m s^\m = 0$. After some straightforward calculations, we arrive at
\begin{equation}
\begin{split}
\label{densityop}
0 &= T^{\m\n}_{\rm sy} \na_{\m} \b_{\n}
	+ T^{\m\n}_{\rm as} \lb \na_{\m} \b_{\n} - 2\o_{\m\n} \rb \\
 &\quad
 	+ S^{\m,\l\n} \na_\m \o_{\l\n}
	+ J^\m (\na_\m \a-F_{\m\n}\b^\n) \,,
\end{split}
\end{equation}
where $T^{\m\n}_{\rm sy/as}$ is the symmetric/antisymmetric part of $T^{\m\n}$.
In the massless case, ${T^\m}_\m=0$, we obtain Eqs.~\eqref{eq:eqml_killing}-\eqref{eq:eqml_alpha} (with $\a_R=\a_L=\a$); In the massive case, we obtain Eqs.~\eqref{eq:eq_killing}-\eqref{eq:eq_alpha}.
Note that one further constraint from Eq.~\eqref{densityop}, $\nabla_{[\m}\o_{\l\n]}=0$, is automatically fulfilled.
\end{appendix}

\bibliography{bibfilespinpolarization}

\end{document}